\documentclass[prc,unsortedaddress,superscriptaddress,a4paper,nofootinbib,%
nobalancelastpage,preprintnumbers,showpacs]{revtex4}%
\usepackage{times}
\usepackage{amssymb,amsbsy,amsmath}
 \usepackage[dvips]{graphicx}
\def\RE{\mathop{\Re{\rm e}}\nolimits}
\def\IM{\mathop{\Im{\rm m}}\nolimits}
\def\vec#1{\boldsymbol{#1}}
\begin{document}
\preprint{\begin{minipage}{3cm}
LPSC 06-36\\
nucl-th/0606050
\end{minipage}
}
\date{\today}
\title{\bf Positivity Constraints on Spin Observables \\
in Exclusive Pseudoscalar Meson Photoproduction}
\author{Xavier Artru}
\email{xavier.artru@ipnl.in2p3.fr}
\affiliation{Institut de Physique Nucl\'eaire de Lyon, CNRS-IN2P3--Universit\'e Lyon-I \\
4, rue Enrico Fermi, 69622 Villeurbanne cedex, France}

\author{Jean-Marc Richard}
\email{jean-marc.richard@lpsc.in2p3.fr}
\affiliation{Laboratoire de Physique Subatomique et Cosmologie\\
Grenoble Universit\'es--CNRS-IN2P3-Ing\'enierie\\
53, avenue des Martyrs, F--38026 Grenoble Cedex, France}

\author{Jacques Soffer}
\email{soffer@cpt.univ-mrs.fr}
\affiliation{Centre de Physique Th\'eorique\protect\footnote{UMR 6207 - Unite mixte de Recherche du CNRS et des Universit\'es Aix-Marseille I,
Aix-Marseille II et de l'Universit\'e du Sud Toulon-Var, Laboratoire affili\'e \`a la FRUMAM.}\\
CNRS, Luminy Case 907, F-13288 Marseille Cedex 09 - France}

\date{\today}
\begin{abstract}
Positivity constraints have proved to be important for spin observables of
exclusive reactions involving polarized initial and final particles.
Attention is focused in this note on the photoproduction of pseudoscalar mesons from spin $1/2$ baryons, more specifically $ \gamma N \to K \Lambda, \gamma N \to K \Sigma$, for which new
experimental data are becoming available.
\end{abstract}
\maketitle
\section{Introduction}\label{se:intro}
Positivity constraints have been widely studied in hadron physics to determine
the allowed domain of physical observables. They can be used to test the consistency
of various available measurements and also the validity of some dynamical assumptions
in theoretical models. This powerful tool has a broad range of applications for the spin
observables in {\it exclusive reactions}, like $\pi N \to \pi N, N N \to N N, N \overline{N} \to
\Lambda \overline{\Lambda}$ and in {\it one-particle inclusive reactions}, like $ pp \to \Lambda X,
ep \to e X, \nu p \to e X $. It also provides constraints for structure functions, parton
distributions, etc.~\cite{aerst}. Here, we concentrate on
the particular reactions $\gamma N \to K Y$, with $Y=\Lambda, \Sigma$, where the incoming
photon beam is polarized, the nucleon target is polarized and the polarization of the
outgoing hyperon is measured.

Many data are available for the reaction $\gamma N \to K Y$ and its analogue without strangeness $\gamma N \to \pi N$. Most of the results are available through the Durham data base \cite{Durham}. Among the recent measurements are those of  the SAPHIR collaboration at Bonn and the LEPS collaboration in Japan. More recently, the photoproduction of a kaon has been studied by the CLAS collaboration at Jlab and the GRAAL collaboration at Grenoble. The results are about to be published.

In this article, it is reminded that the many possible spin observables are not independent, but  constrained by identities and inequalities. They can be established in a systematic way by powerful algebraic methods. The physics content of these identities and inequalities can be revealed by alternative derivations based on the positivity of the density matrix  or by considerations on the norm of the polarization vectors. We also stress that some of the most recent (still preliminary) results seemingly violate these constraints.

Several authors studied how a subset of well-chosen observables enables one to reconstruct the amplitudes to an overall phase. Unavoidably, the question of the  redundancies and compatibility among the various observables is raised, leading to list a number of relations among observables. Our concern is somewhat simpler: we study to which extent a new observable is compatible with the previous data, and which margin is left for the yet-unknown observables whose measurement can be foreseen.

In Sec.~\ref{se:form}, the formalism of the photoproduction reaction $\gamma N\to K Y$ is briefly summarized with some details given in Appendix~\ref{appa}. In Sec.~\ref{se:ineg}, several inequalities relating two or three spin observables are given, and are derived by different methods. Section \ref{se:exp} is devoted to confront the  recent measurements with these model-independent inequalities. Our conclusions are given in Sec.~\ref{se:conc}.
\section{Formalism}\label{se:form}
The formalism of the photoproduction of pseudoscalar mesons has been studied by several authors, see
the pioneer paper by Chew et al.~\cite{Chew:1957tf} and
\cite{GRG,Barker:1975bp,Chiang:1996em}, with particular attention to the  observables which are needed for a full reconstruction of the amplitudes, up to an overall phase.

It is convenient to express the transition matrix $\mathcal{M}$ of reaction $\gamma+N \to K+Y$ in the \emph{transversity basis}: $|\boldsymbol{\pi},\pm\rangle$ and $|\boldsymbol{n},\pm\rangle$  for the initial state and $|\pm\rangle$ for the final state, where
 $\pm$ denotes the transversity of the nucleon or hyperon, i.e., the  projection $\pm1/2$ of its spin along  the normal to the scattering plane, and $\boldsymbol{\pi}$ (resp.\ $\boldsymbol{n}$) a photon state with linear polarization parallel (resp.\ normal)  to the scattering plane.  These states are eigenstates of $\Pi$, the operator  of reflection about the scattering plane and conservation of $\Pi$ is equivalent to parity conservation.

For this parity-conserving exclusive reaction, there are \emph{four} independent transversity amplitudes, which can be chosen as the following matrix elements of the transition operator $\mathcal{M}$
\begin{equation}
\begin{array}{ll}
 a_1=\langle + | \mathcal{M}| \boldsymbol{n}+\rangle~,\quad
 a_2=\langle - | \mathcal{M}| \boldsymbol{n}-\rangle~,\\[2pt]
 a_3=\langle + | \mathcal{M}| \boldsymbol{\pi}-\rangle~,\quad
 a_4=\langle - | \mathcal{M}| \boldsymbol{\pi}+\rangle~,
\end{array}
\label{transvampli}
\end{equation}
while $\langle + | \mathcal{M}| \boldsymbol{n}-\rangle=\langle - | \mathcal{M}| \boldsymbol{n}+\rangle
=\langle + | \mathcal{M}| \boldsymbol{\pi}+\rangle=\langle - | \mathcal{M}| \boldsymbol{\pi}-\rangle=0$.

The complete knowledge of the reaction requires, to an overall phase, the determination of \emph{seven} real functions. On the other hand, one can extract from all the possible experiments \emph{sixteen} different quantities, which are the bilinear products of the four amplitudes.  A well chosen set of observables give access to the amplitudes (to an overall phase) without discrete ambiguities \cite{Barker:1975bp,Chiang:1996em}.

On the experimental side, there are several  redundant observables:
\begin{itemize}\itemsep-2pt
\item the unpolarized differential cross section $I_0$,
\item the linearly-polarized photon asymmetry $\Sigma^\gamma$,
\item the polarized-target asymmetry $A^N$,
\item the  polarization $P^Y$ of the recoiling baryon
\item the baryon depolarization  coefficients $T_i$ and $L_i$ expressing the correlation between the longitudinal or transverse (in the scattering plane) target polarization and the spin of the recoil baryon,
\item the coefficients describing the  transfer of polarization  from a photon beam   to the recoil baryon, in particular $O_i$ for oblique polarization and $C_i$ for  circular polarization,
\item the coefficients $G$, $H$, $E$ and $F$ of
double spin correlations between the photon beam and the nucleon target,
\item
triple correlations coefficients if both the beam and the target are polarized and the hyperon polarization analyzed.
\end{itemize}
In these definitions, the index  $i$ refers to the component in  a frame $\{\hat{x}, \hat{y}, \hat{z}\}$ attached to each particle:
$\hat{y}$, the normal to the scattering plane, is the same for all, and $\hat{z}$ can be chosen along the center-of-mass momentum $\vec{p}$, i.e. $\hat{z}=\vec{p}/p$ (in photoproduction experiments, $\hat{z}=-\vec{p}/p$ is usually chosen for the baryons). For the fermions, it is convenient to use a representation of the spin operators in which $S_y$ is diagonal (by circular permutation of the usual  Pauli matrices). For the photon, the three components $(S^1, S^2, S^3)$ are the Stokes parameters, i.e., $S^3\equiv S_{\ominus}$ (``planarity'') is the polarization along $\hat{x}$, $S^1\equiv S_{\oslash}$ (``obliquity'') is the polarization along $(\hat{x}+\hat{y})/\sqrt{2}$ and $S^2\equiv S_{\odot}$ is the circular polarization, or helicity. With full ($|\vec{S}|=1$) polarization, the  differential cross section can be expressed as
\begin{equation}
{d\sigma \over d\Omega} \left( \vec{S}_\gamma,\vec{S}_N,\vec{S}_Y\right)
= I_0 \,( \lambda\mu|\nu)\,S_\gamma^\lambda \, S_N^\mu \, S_Y^\nu~.
\end{equation}
Here $\lambda, \mu, \nu$, run from 0 to 3, the summation is understood over repeated indices and the polarizations have been promoted to four-vectors with $S^{0} = 1$. The correspondence with the notation found in literature is given in Appendix \ref{appa}, where the observables are listed, and some of their symmetries discussed.

The angular distribution $I_0$ and the product of $I_0$ by a spin observable are bilinear combinations of the four amplitudes. In this paper, the discussion is focused on  eight of the observables listed in Appendix  \ref{appa}, namely
\begin{equation}\label{eq:obs}
\begin{array}{ll}
 I_0=|a_1|^2+|a_2|^2+|a_3|^2+|a_4|^2~,\quad
&I_0 A^N=|a_1|^2-|a_2|^2-|a_3|^2+|a_4|^2~\\[2pt]
I_0 \Sigma^\gamma=|a_1|^2+|a_2|^2-|a_3|^2-|a_4|^2~,\quad
&I_ 0P^Y=|a_1|^2-|a_2|^2+|a_3|^2-|a_4|^2~\\[2pt]
 I_0 C^Y_x=-2 \IM(a_1a_4^*-a_2 a_3^*)~,\quad
&I_0 C^Y_z=2 \RE(a_1a_4^*+a_2 a_3^*)~,\\[2pt]
 I_0 O^Y_x=-2 \RE(a_1a_4^*-a_2 a_3^*)~,\quad
&I_0 O^Y_z=-2 \IM(a_1a_4^*+a_2 a_3^*)~.
\end{array}
\end{equation}
 which are accessible without target polarization, since the analyzing power or target asymmetry $A^N$ is equal to  the transfer of normal polarization from the photon to the hyperon. For the  $C^Y_{x,z}$ and $O^Y_{x,z}$ observables, the expressions given in Eq.~(\ref{eq:obs}) depend on the phase convention among the transversity amplitudes: we follow here Ref.~\cite{Barker:1975bp}. As seen in Appendix \ref{appa}, there are 15 other pairs of observables which are equal or opposite. This means that is is not necessary to measure the reaction with any possible beam and target polarization, except for cross-checks.
\section{Relations among observables}\label{se:ineg}
Each spin observable $\mathcal{O}_i$ is normalized to belong to $[-1,+1]$. However, model-independent inequalities exist among observables, and as a consequence, the allowed domain for a pair  of observables is often smaller than the entire square $[-1,+1]^2$, and similarly for a triple of observables, it is restricted to a sub-domain of the cube $[-1,+1]^3$. The inequalities among observables are of course independent of the choice of amplitudes, e.g., transversity vs.\ helicity amplitudes, and independent of the particular phase conventions which are adopted for these amplitudes. The inequalities do not even depend on the orientation chosen for the $x$- and $z$-axes in the scattering plane, since a rotation about $\hat{y}$ of the $(\hat{x},\hat{y},\hat{z})$ frame attached to a particle only changes the phase of the transversity amplitudes. The inequalities also remain unchanged if we define the spin observables in the laboratory or Breit frame. In these frames, the helicity and the transverse polarization in the scattering plane do not coincide with the center-of-mass ones, but are related by a Wigner rotation about $\hat{y}$.

In Ref.~\cite{GRG}, it is indicated that several pairs of observables $(\mathcal{O}_i,\mathcal{O}_j)$ obey an inequality $
\mathcal{O}_i^2+\mathcal{O}_j^2\le1$,
that restricts the domain to the unit  disk.  Examples are
\begin{equation}\label{disks}
(\Sigma^\gamma)^2+(O^Y_x)^2\le 1~,\quad
(\Sigma^\gamma)^2+(O^Y_z)^2\le 1~,\quad (O^Y_x)^2+(O^Y_z)^2\le 1~.
\end{equation}
For the 7 spin observables given in (\ref{eq:obs}), there are 21 pairs, and 15 of them have this unit-disk constraint.

For triples, there are several examples where the three observables are constrained inside the unit ball, in particular

\begin{subequations}
\centerline{
\begin{minipage}{.40\textwidth}
\begin{eqnarray}
&& (P^Y)^2 + (C^Y_x)^2 + (C^Y_z)^2 \leq 1 ~,\qquad\label{eq:sph1}\\
&& (P^Y)^2 + (O^Y_x)^2 + (O^Y_z)^2 \leq 1 ~,\label{eq:sph2}\\
&& (P^Y)^2 + (C^Y_x)^2 + (O^Y_x)^2 \leq 1 ~,\label{eq:sph3}\\
&& (P^Y)^2 + (C^Y_z)^2 + (O^Y_z)^2 \leq 1 ~,\label{eq:sph4}
\end{eqnarray}
\end{minipage}\hfill \begin{minipage}{.40\textwidth}
\begin{eqnarray}
&& (\Sigma^\gamma)^2 + (C^Y_x)^2 + (C^Y_z)^2 \leq 1 ~,\qquad\label{eq:sph5}\\
&&  (\Sigma^\gamma)^2 + (O^Y_x)^2 + (O^Y_z)^2 \leq 1 ~,\label{eq:sph6}\\
&&  (\Sigma^\gamma)^2 + (C^Y_x)^2 + (O^Y_x)^2 \leq 1 ~,\label{eq:sph7}\\
&&  (\Sigma^\gamma)^2 + (C^Y_z)^2 + (O^Y_z)^2 \leq 1 ~,\label{eq:sph8}
\end{eqnarray}
\end{minipage}}
\end{subequations}

\vspace{.3cm}

\begin{figure}[!h]
\centerline{
\includegraphics[width=.45\textwidth]{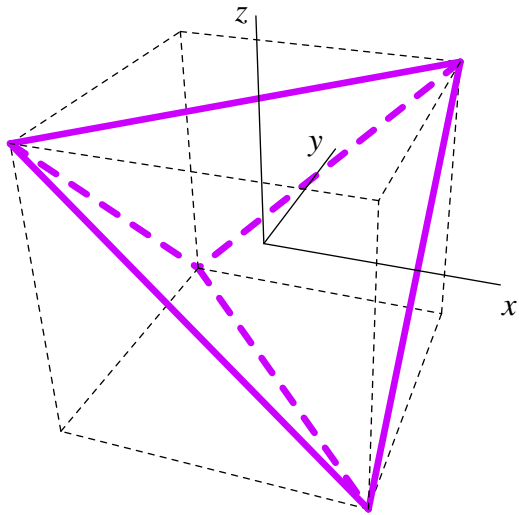}%
}
\caption{(color on line) Tetrahedron domain limited by the inequalities (\protect\ref{tetra}) for the observables  $x=A^N$, $y=P^Y$ and $z=\Sigma^\gamma$.\label{Fig1}}
\end{figure}

By projection, this unit ball  gives a constraint inside the unit disk for any subsystem of two observables.
More interesting is the case where the domain for the three observables is more restricted than the unit cube $[-1,+1]^3$, but \emph{without} restriction for any pair.
For the observables  of $\gamma N\to K Y$, on which data exist, it is known \cite{GRG,Barker:1975bp}
that  $A^N$, $P^Y$ and $\Sigma^\gamma$ fulfill~\cite{GRG}
\begin{equation}\label{tetra}
|A^N-P^Y|  \le 1-\Sigma^\gamma~,\quad
|A^N+P^Y| \le 1+\Sigma^\gamma~.
\end{equation}
These linear relations are simply obtained from the positivity of the pure transversity cross sections $|a_i|^2$ in (\ref{eq:obs}). They are also found for spin observables of inclusive reactions \cite{aerst,Soffer:2003qj}. The domain corresponds to a tetrahedron schematically drawn in Fig.~\ref{Fig1}. Notice that its volume is only 1/3 of the volume of the entire cube, while its projection is the entire $[-1,+1]^2$ square on any face.

There are several methods to establish the above inequalities. The most straightforward consists of plotting dummy observables from amplitudes whose real and imaginary parts are generated randomly. The plots  clearly indicate whether the full square $[-1,+1]^2$ or  cube $[-1,+1]^3$ is entirely scanned, or  the domain is limited by a circle, a triangle, a sphere, etc. Then for these pairs or triples, the  observed inequalities can be derived from the explicit expressions such as (\ref{eq:obs}). With the transversity amplitudes, $A^N$, $P^Y$ and $\Sigma^\gamma$ have simple expressions, and the tetrahedron constraint is easily seen, while it is less obvious with other choices, e.g., helicity amplitudes, for which, in turn, other inequalities become  easier.

The proofs remain at the level of elementary calculus. For instance, Eq.~(\ref{eq:sph7}) can be obtained from the inequality $(\vec{V}_1-\vec{V}_2)^2\le (|\vec{V}_1|+|\vec{V}_2|)^2$ applied to the vectors $\vec{V}_1=\{|a_1|^2-|a_4|^2, 2\RE(a_1a_4^*),2\IM(a_1a_4^*)\}$ and $\vec{V}_2=\{|a_3|^2-|a_2|^2,2\RE(a_2a_3^*),2\IM(a_2a_3^*)\}$, with normalization  $|V_1|=|a_1|^2+|a_4|^2$ and $|V_2|=|a_2|^2+|a_3|^2$.  A slight variant, e.g.\ for Eq.~(\ref{eq:sph8}), consists of checking that the four-vectors $(1\pm A^N;\,P^Y\pm \Sigma^\gamma,\,O^Y_z\pm C^Y_x,\,O^Y_x\pm C^Y_z)$ are light-like, and hence, since their time components  $1\pm A^N$ are positive, their sum is time-like, in the same way as two photons combine into a massive state in elementary kinematics.

Another possible starting point is the existence of identities among observables.
A consequence already stressed in the literature \cite{GRG,Barker:1975bp,Chiang:1996em} is
that if sixteen measurements can be expressed
in terms of seven real functions, they cannot be independent and
must be constrained by a number of  relations.
Several years ago, G.~Goldstein et al., following earlier work by Fr{\o}yland and Worden \cite{GRG}, derived nine quadratic relations among these parameters.  
The method of Fierz transformation \cite{Chiang:1996em} provides a systematic list of identities. Most of them, however, involve many observables, and hence it is not obvious to convert these identities into inequalities constraining the few observables for which data exist.

Another powerful tool is provided by imposing the positivity of the density matrix for the direct or crossed channels. If the $\gamma+N\to \Lambda+X$ reaction is viewed as crossed from $\gamma+N+\overline{\Lambda}\to X$, the density matrix has dimension $8\times8$, since each of the incoming particle has two spin degrees of freedom.  This is reduced to $4\times4$ if the parity of $X$ is identified. In our case, since $X=K$ has spin 0, this $4\times4$ density matrix has rank 1, and hence each $2\times 2$ subdeterminant vanishes. This gives 36 quadratic relations which are those listed in \cite{Chiang:1996em} or linear combinations of them. Only 9 of them can be independent, since the observables depend on 7 independent real parameters. However it does not mean that the 9 independent relations alone induce the 27 other ones. For instance, the vanishing of the nine sub-determinants formed by the intersection of two consecutive columns with two consecutive lines induces the vanishing of all the other determinants only if the $2^{\text{nd}}$ and $3^{\text{rd}}$ columns are non-zero. This shows that the 2$\times$2 sub-determinants are related in a non-linear way and 27 of them cannot be expressed from the remaining 9 ones without discrete ambiguities.  

More explicitly, following the method of \cite{Artru:2004jx}, the density matrix of $\gamma+N+\overline{\Lambda}\to K$ can be defined as 
\begin{equation}
\left\langle \boldsymbol{e}', s'_N, s'_Y \left| R \right| \boldsymbol{e}, s_N, s_Y \right\rangle
= C \ \langle  \boldsymbol{e}', s'_N | \mathcal{M}^\dagger | s'_Y  \rangle
\langle s_Y   | \mathcal{M} | \boldsymbol{e}, s_N \rangle~,
\label{Rphotprod}
\end{equation}
where $C$ is a normalization coefficient. By construction $R$ is semi-positive definite and of rank 1. In terms of the observables for $\gamma+N\to K + \Lambda$, 
\begin{equation}
R = (\lambda\mu|\nu) \ \sigma^\lambda_\gamma \otimes
\sigma^\mu_N \otimes \left[\sigma^\nu_Y\right]^t~.
\label{decompose-R}
\end{equation}
Note in (\ref{Rphotprod}) the crossing $|s_Y\rangle \leftrightarrow \langle s_Y|$  of the hyperon and the corresponding transposition of $\sigma^\nu_Y$ in (\ref{decompose-R}).
Table \ref{tab-app} of Appendix \ref{appa} gives the matrix elements of R. 
For instance, from the vanishing of its co-diagonal 2$\times$2 sub-determinants we obtain
\begin{subequations}
\begin{eqnarray}
(1\pm A^N)^2 &=& (P^\gamma\pm \Sigma)^2 + (O^Y_z \pm  C^Y_x)^2 + (C^Y_z \mp O^Y_x)^2
\label{quadratic-1}~,	\\
(1\pm P^Y)^2 &=& (A^N\pm \Sigma)^2 + (G \mp F)^2 + (E \pm H)^2
\label{quadratic-2}~, \\
(1\pm \Sigma^\gamma)^2 &=& (P^Y\pm A)^2 + (L_z \pm T_x)^2 + (L_x\mp T_z)^2~.
\label{quadratic-3}
\end{eqnarray}
\end{subequations}

Complex identities, i.e., pairs of real identities, which do not contain $P^Y$, $A^N$ nor $\Sigma^\gamma$ can be obtained from the 2$\times$2 determinants  made only of non-diagonal elements of Table 1, for instance
\begin{equation}
(E+iG)^2+(F+iH)^2+(O_z-iC_z)^2+(O_x-iC_x)^2=0~.
\label{quadratic-5}
\end{equation}

Many identities can be listed \cite{Chiang:1996em}, but they are related by symmetry rules, in particular
\begin{itemize}
\item
rotation in the $(x,y)$ plane, in particular the substitution $x\to y$, $y\to -x$~,
\item
permutation of the $\lambda,\mu,\nu$ indices, i.e., of the three particles with spin, except for the transposition of 
$\sigma^\nu_Y$. For instance, Eqs.~(\ref{quadratic-1}) and (\ref{quadratic-3}) are related by such a transformation,
\item
$\pi/2$ rotation in the (1,3) plane: 1$\to$3, 3$\to -1$ for the three particles simultaneously. This is not compatible with the parity rules such as $(32|2)=0$ or the equivalences of Eqs.~(\ref{eq:obs-def}). However, relaxing parity temporarily, one can apply the (1,3) rotation to Table 1, then enforce parity conservation to reduce the new matrix elements.  For instance, this transformation applied to  Eq.(\ref{quadratic-3}) leads to 
\begin{equation}
(1+L_z)^2 = (E+C_z)^2 + (\Sigma+T_x)^2 + (G+O_z)^2~,
\label{rot1-3-ex}
\end{equation}
\item
substitution $0 \to i1$, $1 \to i0$ for all the particles, corresponding to an imaginary Lorentz transformation of the $\sigma_\mu$'s in the (0,1) plane. Equation (\ref{quadratic-3}), for instance, is transformed into
\begin{equation}
(1-T_x)^2 = (F+C_x)^2 + (\Sigma-L_z)^2 + (H-O_x)^2~.
\label{rot1-0-ex}
\end{equation}
\end{itemize}

As an example of application, Eq.~(\ref{quadratic-1}) implies 
\begin{equation}
(1\pm A^N)^2\ge (P^Y\mp \Sigma^\gamma)^2~,
\end{equation}
which is equivalent to (\ref{tetra}).

Note also that some  inequalities just follow from the definition of the observables.   Equation (\ref{eq:sph1}), for instance,  expresses the usual bound on the three components of the polarization vector of the recoil baryon, when the photon is, say, right-handed.
Similarly, considering a photon with oblique linear polarization, one obtains (\ref{eq:sph2})
of which the last inequality in (\ref{disks}) is a consequence.

The first of the inequalities (\ref{disks}) is implied by  the  more constraining inequality (\ref{eq:sph7})
which can be understood as follows: if the reverse reaction is performed with an hyperon fully polarized along the $x$ axis, then the outgoing photon receives a polarization whose components are precisely $C^Y_x$, $\Sigma^\gamma$, $O^Y_x$, leading to (\ref{disks}).
Similarly, for the second inequality of (\ref{disks}), one can add $(C_z ^Y)^2$.

For demonstrating (\ref{eq:sph6}), it is conceivable to rotate the axis in the $(x,z)$ plane, say $(x,z)\to (u,v)$ such that
$C^Y_u$ is maximal and $C^Y_v$ vanishes. Then the reverse reaction can be envisaged, with a polarization along the $u$ axis for the hyperon. If the various components of the polarization of the outgoing photon are considered, an inequality
\begin{equation}\label{sphere-bis}
(C^Y_u)^2+(\Sigma^\gamma)^2+(O^Y_u)^2\le 1~,
\end{equation}
is deduced from which (\ref{eq:sph6}) follows by neglecting the first term and noticing that
\begin{equation}\label{sphere-bis1}
(O^Y_u)^2=(O^Y_x)^2+(O^Y_z)^2~.
\end{equation}
\section{Confronting data}\label{se:exp}
In a recent paper, the LEPS collaboration published results for the photoproduction reaction
$\gamma n \to K^+\Sigma^-$  for incident photon energy from 1.5 to 2.4 GeV \cite{Kohri:2006yx}. It is remarkable that $\Sigma^\gamma$ is close to $\Sigma^\gamma=1$ in a wide range of energies for centre-of-mass angle such that $\cos(\theta_{cm})>0.6$.  Equation (\ref{tetra}) implies that $A^N\simeq P^Y$.

The CLAS collaboration has studied the reaction $\gamma p \to K^+ Y$ measured for center-of-mass energies $W$ between 1.6 and 2.53 GeV and for $ -0.85 < \cos\theta_K^{c.m} < +0.95$ \cite{Bradford:2005pt,nabb,brad04}. In addition to the differential cross-section, three spin observables are measured,
namely $P^Y$ and the double
correlation parameters $C^Y_x$ and $C^Y_z$ between the circularly polarized photon and the recoil
baryon spin along the directions $\hat x$ and $\hat z$ in the scattering plane. In these preliminary data, it is  observed
that some values of $P_{\Lambda}$ are very large, for example $P_{\Lambda}=-0.73$  at $W=1.73$ GeV and $\cos\theta_K^{c.m} = +0.30$. From Eq.~(\ref{eq:sph1}), this result leads to $(C^{\Lambda}_x)^2 + (C^{\Lambda}_z)^2 \leq 0.5$,
which seems inconsistent with the value $C^{\Lambda}_z \sim 1$, reported in Ref.~\cite{brad04}. Of course, we have  to
wait for the final data before drawing any definite conclusion, but we urge the CLAS Collaboration
to make sure that the above constraint is indeed fulfilled.

Very recently, Schumacher \cite{Schumacher:2006ii} stressed that in the CLAS data
\begin{equation}
(P^Y)^2+(C^Y_x)^2+(C^Y_z)^2\simeq 1~.
\end{equation}
It follows from  (\ref{quadratic-1}) that
\begin{equation}
(A^N)^2 \simeq ( \Sigma^\gamma)^2+(O^Y_x )^2 + (O^Y_z)^2~,
\end{equation}
should also be verified.

The GRAAL collaboration has measured the beam asymmetry $\Sigma^\gamma$ for $\pi^0$ photoproduction \cite{Bartalini:2005wx}. Large positive values are often found, for instance $\Sigma^\gamma=0.990\pm0.069$ at incident energy $E_\gamma=1344\,$MeV and center-of-mass angle $\theta_{\text{cm}}=45.1^\circ$. From Eq.~(\ref{tetra}), this implies the non trivial equality $A^N\simeq P^Y$ of analyzing power  and recoil nucleon polarization. The same collaboration
is about to publish its results on three spin observables for kaon photoproduction: the hyperon polarization $P^Y$, the beam asymmetry $\Sigma^\gamma$ and the correlation coefficients $O^Y_x$ and $O^Y_z$ between the  photon oblique linear polarization and the polarization of the hyperon \cite{Graal}. The  inequalities (\ref{disks})  are seemingly far from saturation.
\section{Outlook}\label{se:conc}
The new data on photoproduction of pseudoscalar mesons, in particular $\gamma N\to K Y$ include several spin observables, which can discriminate among the different models. Before undertaking a phenomenological analysis, it is crucial to check that the various spin observables are compatible. The ultimate criterion would consist of obtaining a consistent set of amplitudes. A more immediate test is to check whether all the possible model-independent inequalities are fulfilled by the data.

A similar approach  can be applied to other exclusive reactions for which several spins are measured, in particular the strangeness-exchange reaction  $\bar{p}p\to \overline{\Lambda}\Lambda$ using a polarized target. This study has been done at the LEAR facility of CERN and will probably be resumed at higher energy at the forthcoming FAIR complex at Darmstadt. With minor changes, the same formalism of inequalities also holds  for the observables of inclusive reactions, and for the spin-dependent structure functions and parton distributions \cite{aerst}.

At first sight, deriving inequalities among observables is a mere algebraic exercise applied to the bilinear relations expressing these observables in terms of the independent amplitudes, once the symmetry constraints have been imposed.  In fact, these inequalities reflect the positivity of the density matrix for the initial and final states of the reaction in the direct and crossed channels.  For any reaction, the spin state in the initial state, factorizable or entangled, undergoes a quantum transformation, and is thus submitted to the general rules on the transmission of information in elementary quantum processes.

\appendix
\section{Observables}\label{appa}
The definition of $\Sigma^\gamma$, $A^N$, ... $C^Y_x$ is taken from \cite{Fasano:1992es}. They differ in sign with \cite{Barker:1975bp} concerning $L_x$, $G$, $E$, $C^Y_x$ and $C^Y_z$.  
\begin{subequations}\label{eq:obs-def}
\centerline{
\begin{minipage}{.50\textwidth}
\begin{eqnarray}
 ( 00|0 ) &=& - ( 33|3 ) = 1 \\
 \langle \ominus \rangle = - \langle yy' \rangle = ( 30|0 ) &=& - ( 03|3 ) = -\Sigma^\gamma \\
 \langle y \rangle = - \langle \ominus y' \rangle = ( 03|0 ) &=& - ( 30|3 ) = + A^N \\
 \langle y' \rangle = - \langle \ominus y \rangle = ( 00|3 ) &=& - ( 33|0 ) = + P^Y \\
 \langle zz' \rangle = ( 01|1 ) &=& - ( 32|2 ) = + L_z \\
 \langle zx' \rangle = ( 01|2 ) &=& + ( 32|1 ) = + L_x \\
 \langle xz' \rangle = ( 02|1 ) &=& + ( 31|2 ) = + T_z \\
 \langle xx' \rangle = ( 02|2 ) &=& - ( 31|1 ) = + T_x
 \end{eqnarray}
\end{minipage}\hfill \begin{minipage}{.48\textwidth}
\begin{eqnarray}
 \langle \oslash z \rangle = ( 11|0 ) &=& + ( 22|3 ) = - G \\
 \langle \oslash x \rangle = ( 12|0 ) &=& - ( 21|3 ) = - H \\
 \langle \odot z \rangle = ( 21|0 ) &=& - ( 12|3 ) = + E \\
 \langle \odot x \rangle = ( 22|0 ) &=& + ( 11|3 ) = + F \\
 \langle \oslash z' \rangle = ( 10|1 ) &=& - ( 23|2 ) = - O_z^Y \\
 \langle \oslash x' \rangle = ( 10|2 ) &=& + ( 23|1 ) = - O_x^Y \\
 \langle \odot z' \rangle = ( 20|1 ) &=& + ( 13|2 ) = + C_z^Y \\
 \langle \odot x' \rangle = ( 20|2 ) &=& - ( 13|1 ) = + C_x^Y
\end{eqnarray}
\end{minipage} }
\par\vspace{.3cm}\noindent
\end{subequations}
The symbol $\langle\oslash x'\rangle$, for instance, is an intuitive notation for the correlation between the oblique polarization of the photon (at $+\pi/4$) and the polarization towards $\hat{x}$ of the final baryon.

In the transversity basis, owing to (\ref{transvampli}), $R=R^+ \oplus R^-$, where $R^-$ acts in the subspace spanned by
\begin{equation}
\left|\boldsymbol{\pi}+- \right\rangle~,\quad
\left|\boldsymbol{\pi}-+  \right\rangle~,\quad
\left|\boldsymbol{n}++\right\rangle \quad\text{and}\quad
\left|\boldsymbol{n}--\right\rangle~,
\end{equation}
and  where $R^+$, acting on the complementary subspace, vanishes identically. 

The matrix $R^-$ is given by Table~\ref{tab-app},  and 
$( 0+3\,,\,1-i2\,|\,1+i2 )$, for instance, is a compact notation for 
\begin{equation}
( 01|1 ) + ( 31|1 ) -i ( 02|1 ) -i ( 32|1 )+i ( 01|2 ) +i ( 31|2 ) + ( 02|2 ) + ( 32|2 )~.
\label{compact}
\end{equation}

\begin{table}[!hb]
\caption{\label{tab-app}Sub-matrix $R^-$ of the density matrix $R$.}
$$
             \begin{array}{|c|c|c|c|c|}
		\hline 
 & \pi+- & \pi-+ & n++ & n-- \\     \hline 
\pi+- & ( 0+3\,,\,0+3\,|\,0-3 ) & ( 0+3\,,\,1-i2\,|\,1-i2 ) & ( 1-i2\,,\,0+3\,|\,1-i2 ) & ( 1-i2\,,\,1-i2\,|\,0-3 )
\\ \hline 
\pi-+ & ( 0+3\,,\,1+i2\,|\,1+i2 ) & ( 0+3\,,\,0-3\,|\,0+3 ) & ( 1-i2\,,\,1+i2\,|\,0+3 ) & ( 1-i2\,,\,0-3\,|\,1+i2 )
\\ \hline 
n++ & ( 1+i2\,,\,0+3\,|\,1+i2 ) & ( 1+i2\,,\,1-i2\,|\,0+3 ) & ( 0-3\,,\,0+3\,|\,0+3 ) & ( 0-3\,,\,1-i2\,|\,1+i2 )
\\\hline 		
n-- & ( 1+i2\,,\,1+i2\,|\,0-3 ) & ( 1+i2\,,\,0-3\,|\,1-i2 ) & ( 0-3\,,\,1+i2\,|\,1-i2 ) & ( 0-3\,,\,0-3\,|\,0-3 )
\\\hline 		
                   \end{array}
$$
\end{table}

The 16 equivalences in Eqs.(\ref{eq:obs-def}a-p) reflect  the  invariance of $(\lambda,\mu|\nu)$  under the  substitutions: $0 \to -3$, $3 \to -0$, $1\to i2$, 2$\to -i1$ for $\lambda$ and $\mu$ and 0$\to -3$, 3$\to -0$, 1$\to -i2$, 2$\to i1$ for $\nu$. It comes from the vanishing of amplitudes between an even number of particles with negative $\sigma_3$, expressed by 
\begin{equation}
\sigma_3 \otimes \sigma_3 \otimes \sigma_3 \ R = R\ \sigma_3 \otimes \sigma_3 \otimes \sigma_3 = -R~.
\label{equivalence}
\end{equation}
Using these equivalences, one can simplify $R$ by replacing $m\pm n$ by $m$ ($m=0$ or 1) {\it once} in each box of Table \ref{tab-app}. The result is equal to $R/2$.

\begin{acknowledgments}
Informative discussions with Reinhard  Schumacher (CLAS), Annick Lleres, Jean-Paul Bocquet and Dominique Rebreyend (GRAAL),  Takashi Nakano (LEPS), and   Oleg Teryaev  are gratefully acknowledged, as well as the comments by Muhammad Asghar.
\end{acknowledgments}

\end{document}